\documentclass[aps,prl,twocolumn,preprintnumbers,showpacs]{revtex4}
\usepackage{graphicx}
\usepackage{soul}
\usepackage{color}
\usepackage[normalem]{ulem}  

\begin{document}

\title{The Neutron Star Mass-Radius Relation and the \\
Equation of State of Dense Matter }

\author{Andrew W. Steiner$^{1,2}$}
\author{James M. Lattimer$^{3}$}
\author{Edward F. Brown$^{2}$}

\affiliation{$^{1}$Joint Institute for Nuclear Astrophysics, 
National Superconducting} 
\affiliation{Cyclotron Laboratory, and the Department of Physics 
and Astronomy,} 
\affiliation{Michigan State University, East Lansing, MI 48824} 
\affiliation{$^{2}$Institute for Nuclear Theory, University of Washington, 
Seattle, WA 98195} 
\affiliation{$^{3}$Department of Physics \& Astronomy, Stony Brook University, 
Stony Brook, NY 11794} 

\begin{abstract} 
The equation of state (EOS) of dense matter has been a long-sought
goal of nuclear physics. Equations of state generate unique mass
versus radius ($M\textrm{--}R$) relations for neutron stars, the
ultra-dense remnants of stellar evolution. In this work, we determine
the neutron star mass-radius relation and, based on recent
observations of both transiently accreting and bursting sources, we
show that the radius of a 1.4 solar mass neutron star lies between
10.4 and 12.9 km, independent of assumptions about the composition of
the core. We show, for the first time, that these constraints remain
valid upon removal from our sample of the most extreme transient
sources or of the entire set of bursting sources; our constraints also
apply even if deconfined quark matter exists in the neutron star core.
Our results significantly constrain the dense matter EOS and are,
furthermore, consistent with constraints from both heavy-ion
collisions and theoretical studies of neutron matter. We predict a
relatively weak dependence of the symmetry energy on the density and a
value for the neutron skin thickness of lead which is less than 0.20
fm, results that are testable in forthcoming experiments.
\end{abstract}

\preprint{INT-PUB-12-028}

\maketitle

The masses of several neutron stars have been precisely measured using
pulsar timing~\cite{Lorimer98}; simultaneous mass and radius
measurements, however, are considerably less certain. The leading candidates for
such measurements are bursting neutron stars that show photospheric
radius expansion (PRE)~\cite{vanParadijs79} (for a review, see Lewin
et al. 1993~\cite{Lewin93}) and transiently accreting neutron stars in
quiescence~\cite{Rutledge99}. 

Observations have already begun to determine the universal $M\textrm{--}R$
relation for neutron stars and to place strong constraints on the EOS
of dense matter. Previously derived
constraints~\cite{Read09,Ozel10,Steiner10,Steiner12} have several
limitations, including the use of fixed parametrizations for the EOS.
Those works did not show their results to be independent of their
parametrizations, including the possibility of deconfined quark
matter. Neither did they address the full set of sources analyzed
here. \"{O}zel et al.~\cite{Ozel10} considered only PRE sources, but
these may be subject to considerable systematic
errors~\cite{Steiner10,Boutloukos10,Suleimanov11}. Steiner et
al.~\cite{Steiner10} considered both types of sources but used a
smaller data set. Here we use eight neutron stars: four which produced
PRE X-ray bursts (4U~1608--522~\cite{Guver10},
KS~1731--260~\cite{Ozel12}, EXO~1745--248~\cite{Ozel09}, and
4U~1820--30~\cite{Guver10b}) and four in quiescent low-mass X-ray
binaries (qLMXBs) in the globular clusters M13~\cite{Webb07}, $\omega$
Cen~\cite{Webb07}, 47 Tuc~\cite{Heinke06} and NGC
6397~\cite{Guillot10}. Refs.~\cite{Hebeler10,Steiner12} took advantage
of modern predictions for pure neutron matter near the saturation
density to constrain the $M\textrm{--}R$ relation, but ignored systematic
uncertainties associated with the observations. In this paper, we
demonstrate that the inferred $M\textrm{--}R$ constraints are insensitive to the
removal of either all the PRE burst sources or the most
extreme qLMXB sources. 

At the lowest energy densities ($\le15$~MeV/fm$^3$ or $\lesssim 3 \times
10^{13}$ g/cm$^3$) the pressure-density relation is
well-understood~\cite{Haensel06}. Between 15 and 200--300 MeV/fm$^3$,
the EOS is well-described by four parameters, the incompressibility,
the skewness, the magnitude of the symmetry energy ($S_v$), and the
parameter describing the density derivative of the symmetry energy
($L$), all evaluated at the nuclear saturation density (approximately
150 MeV fm$^{-3}$). These parameters are constrained to varying
degrees by experimental data~\cite{Lattimer12}, including nuclear
masses~\cite{Kortelainen10}, neutron skin thicknesses~\cite{Chen10},
giant dipole resonances and dipole
polarizabilities~\cite{Trippa08,Tamii11,Piekarewicz12}, and heavy-ion
collisions~\cite{Tsang09}. High-density matter is constrained by (i)
causality (the speed of sound must not exceed the speed of light),
(ii) hydrodynamical stability, and (iii) having a sufficient maximum mass (it must be
greater than the largest well-determined neutron star mass,  $1.97\pm0.04\,\mathrm{M}_{\odot}$ for PSR J1614-2230~\cite{Demorest10}). In addition, we impose the constraint that implied neutron star
masses cannot be less than what is achievable in supernova, about 0.8
M$_{\odot}$. 

The low-density part of all the EOS parametrizations (except for
strange quark stars) are described as in Steiner et
al.~\cite{Steiner10}. All of the parameters in the low-density EOS, as
well as the parameters in the high-density EOS models, are described
with uniform prior distributions. Our fiducial EOS model (A)
parametrizes the high-density EOS as a set of two piecewise continuous
power laws defining pressure $P=\varepsilon^{1+1/n}$ as a function of
energy density $\varepsilon$. Model A has four high-density
parameters: the transition energy density between the low-density EOS
and the first polytrope, the transition energy density between the
first and second polytrope, and the two polytropic indices, $n_1$ and
$n_2$. We also employ an EOS model (B) which is similar to model A,
except that the exponents in the two polytropes are parametrized with
uniform priors in $\Gamma_i$ instead of $n_i$. A third EOS model (C),
parametrizes the EOS at high densities with a uniform prior in the
pressure at four fixed energy densities, 400, 600, 1000, and 1400
MeV/fm$^3$. The low-density EOS is used up to energy densities of 200
MeV/fm$^3$ and the EOS is assumed to be linear between that point and
400 MeV/fm$^3$. The linear relation between energy densities of 1000
and 1400 MeV/fm$^3$ is extrapolated to higher energy densities when
necessary. All parameters are chosen with large enough ranges to
ensure the results do not change significantly when the range is
increased.

A fourth EOS model (D) assumes that matter is a polytrope at
intermediate densities and quark matter at high densities. The
pressure of quark matter is described by~\cite{Alford05}
\begin{equation}
P = \frac{3 a_4}{4 \pi^2} \mu^4 - \frac{3 a_2}{4 \pi^2} \mu^2 -B  \, ,
\end{equation}
where $\mu$ is the quark chemical potential. The quantity $B$ is the
bag constant and simulates confinement. The parameter $a_4$ describes
corrections to the leading coefficient from a non-interacting Fermi
gas (for which $a_4=1$). Corrections from perturbative quantum
chromodynamics at high density suggest $0.6 < a_4 < 1$. The parameter
$a_2$ approximately describes corrections from the finite strange
quark mass $m_s$ and the quark superfluid gap $\Delta$ and is given by
$a_2=m_s^2 - 4 \Delta^2$ (see also Ref.~\cite{Steiner12}). In order to
include the effects of a possible mixed phase in a model-independent
way, a polytrope is added in between the low-density EOS and quark
matter. In this model, there are five parameters in total: the
transition energy density between low densities and the polytrope, the
polytropic index, the transition energy density between the polytrope
and quark matter (used to fix the value of $B$), $a_2$, and $a_4$. To
describe bare strange quark stars, Model E applies Equation 1 at all
densities, with neither a low-density EOS nor an intermediate
polytrope. Model E thus has three parameters, $a_2$, $a_4$, and $B$.

Our baseline data set includes all eight astrophysical sources,
interprets the PRE burst sources with an extended photosphere,
and assumes the same distribution of color correction factors and
distances as in Steiner, et al.~\cite{Steiner10}. We consider several
modifications to the baseline scenario. Suleimanov et
al.~\cite{Suleimanov11} have suggested that the X-ray spectra for PRE
sources are affected by accretion and the eclipse of the neutron star
by the disk. This affects the normalization at late times and we take
this into account by increasing the color correction factor $f_C$,
taking $1.45<f_C<1.8$ (modification I). Alternatively, Boutloukos et
al.~\cite{Boutloukos10} have suggested that the color correction
factors are, instead, smaller as a result of magnetic confinement of
the X-ray burst; this is considered in modification II in which we
assume $1<f_C<1.35$. Some previous works assumed the photosphere of
PRE bursts is coincident with the neutron star surface~\cite{Ozel10},
and we also consider this scenario (III). Finally, we test the
sensitivity of our results to the removal of any one source or class
of sources by removing X7 (IV) or M13 (V), and by removing
all PRE sources (VI). To summarize, the standard models we examine
include the baseline case (A), three variations of the EOS (B--D),
and six other modifications of the baseline model varying the included
data or its interpretation (A I--A VI). In addition, we also
examine several more speculative scenarios which are described below.

We use the Bayesian method of Ref.~\cite{Steiner10}, using marginal
estimation to determine the posterior probability densities of
quantities of interest. The marginal estimation integrals are
performed using Markov Chain Monte Carlo.

\begin{table}
\begin{tabular}{llrrrr}
\hline
EOS Model & Data modifications & $R_{95\%>}$ & $R_{68\%>}$ & $R_{68\%<}$ 
& $R_{95\%<}$ \\
& & \multicolumn{4}{c}{(km)} \\
\hline
\multicolumn{6}{c}{Variations in the EOS model}\\
\hline
A & -        & 11.18 & 11.49 & 12.07 & 12.33 \\
B & -        & 11.23 & 11.53 & 12.17 & 12.45 \\
C & -        & 10.63 & 10.88 & 11.45 & 11.83 \\
D & -        & 11.44 & 11.69 & 12.27 & 12.54 \\
\hline
\multicolumn{6}{c}{Variations in the data interpretation}\\
\hline
A & I & 11.82 & 12.07 & 12.62 & 12.89 \\
A & II & 10.42 & 10.58 & 11.09 & 11.61 \\
A & III & 10.74 & 10.93 & 11.46 & 11.72 \\
A & IV & 10.87 & 11.19 & 11.81 & 12.13 \\
A & V  & 10.94 & 11.25 & 11.88 & 12.22 \\
A & VI  & 11.23 & 11.56 & 12.23 & 12.49 \\
\multicolumn{2}{l}{Global limits} & 10.42 & 10.58 & 12.62 & 12.89\\
\hline
\multicolumn{6}{c}{More speculative scenarios}\\
\hline
C & II & 9.17 & 9.34 & 9.78 & 10.07 \\
A & VII & 12.14 & 12.29 & 12.63 & 12.81 \\
E & - & 10.19 & 10.64 & 11.57 & 12.01 \\
A & VIII & 12.35 & 12.83 & 13.61 & 13.92 \\
\hline
\end{tabular}
\caption{Limits for the radius of a 1.4 solar mass neutron star 
for all of the models considered in this work. Model A and
the assumption $1.33 < f_C < 1.47$ for the PRE sources is 
assumed unless specified otherwise.}
\label{tab:radii}
\end{table}

In our baseline analysis, we find strong constraints on the $M\textrm{--}R$ curve
and on the dense matter EOS: the radius of a $1.4~\mathrm{M}_{\odot}$
neutron star is between 11.2 and 12.3 km (95\% confidence). The
permissible radius range encompassing all variations of the EOS and
interpretations of the astrophysical data, but not including the more
speculative scenarios, is 10.4--12.9 km (95\% confidence), only
moderately larger than the baseline result. The 68\% and 95\%
confidence ranges are displayed in the upper and middle portions of
Table~\ref{tab:radii}. We determine the $M\textrm{--}R$ relation for a range of
neutron star masses (see Figure 1). We also determine
the 68\% and 95\% confidence intervals of the EOS of dense matter
(Figure 2). The estimated uncertainty of the pressure is approximately
30--50\% at all densities achievable in neutron star interiors. In
addition, the posterior distributions of the central energy density of
the maximum mass star imply that the highest central density is $\sim
1200$ MeV/fm$^{3}$~\cite{Lattimer05}.

\begin{figure}
  \includegraphics[width=3.4in]{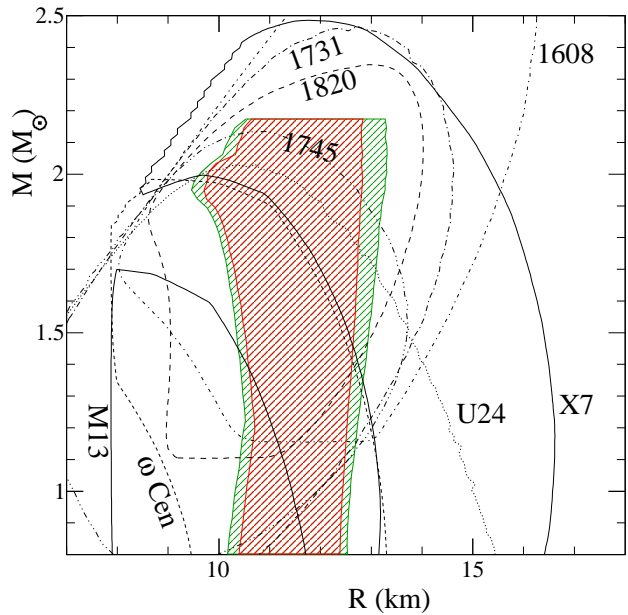}
  \caption{A comparison of the predicted $M\textrm{--}R$ relation with the
    observations. The shaded regions outline the 68\% and 95\%
    confidences for the $M\textrm{--}R$ relation;
    these include variations in the EOS model and the modifications to
    the data set (see Table~\ref{tab:radii}) but not the more
    speculative scenarios. The lines give the 95\% confidence regions
    for the eight neutron stars in our data set.}
  \label{fig:mvsr}
\end{figure}

\begin{figure}
  \includegraphics[width=3.4in]{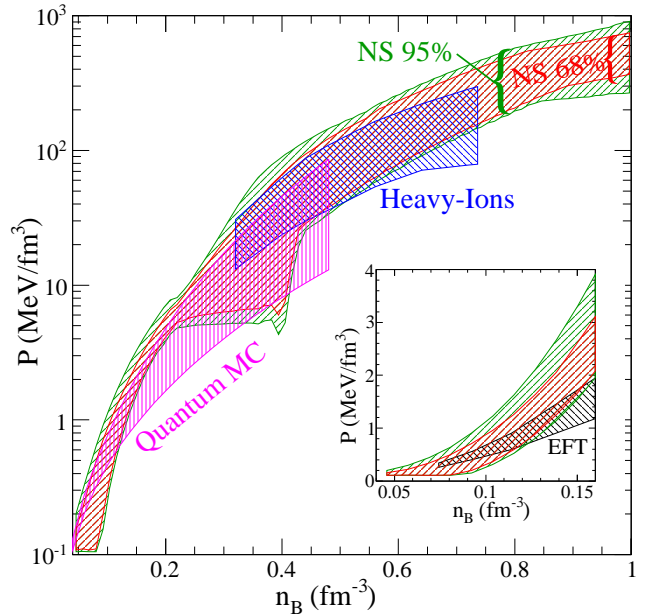}
  \caption{The predicted pressure as a function of baryon density of
    neutron-star matter as obtained from astrophysical observations.
    The region labeled ``NS 68\%'' gives the 68\% confidence limits
    and the region labeled ``NS 95 \%'' gives the 95\% confidence
    limits. Results for neutron-star matter from effective field
    theory~\cite{Hebeler10} (see inset), from quantum Monte
    Carlo~\cite{Gandolfi12}, and from constraints inferred from
    heavy-ion collisions~\cite{Danielewicz02} are also shown for
    comparison.}
  \label{fig:neut}
\end{figure}

Producing significantly different neutron star radii requires
extreme assumptions regarding the EOS and the data. We now consider
more speculative scenarios, which are presented in the bottom portion
of Table~\ref{tab:radii} (see also Figure~3). To achieve significantly
smaller radii, we must assume both that the color correction factor is
anomalously small ($\le1.3$) for all of the PRE sources and that the
EOS has strong phase transitions (model C). In this case, we get radii
as small as 9 km. Increasing the maximum mass
constraint, as would be the case if the estimated most-likely mass of
the pulsar B1957+20 is 2.4 M$_{\odot}$~\cite{vanKerkwijk11}, slightly
increases radii (modification VII). We obtain even larger radii
if we add the long PRE burst source 4U 1724--307~\cite{Suleimanov11}
and further assume, as suggested in Suleimanov et
al.~\cite{Suleimanov11} (modification VIII), that the short PRE bursts
and the qLMXBs M13 and $\omega$ Cen not be considered because of
modifications to their spectra due to accretion. This
scenario cannot yet explain, however, why short PRE burst cooling tails are
observed to have constant normalizations.

\begin{figure}
  \includegraphics[width=3.4in]{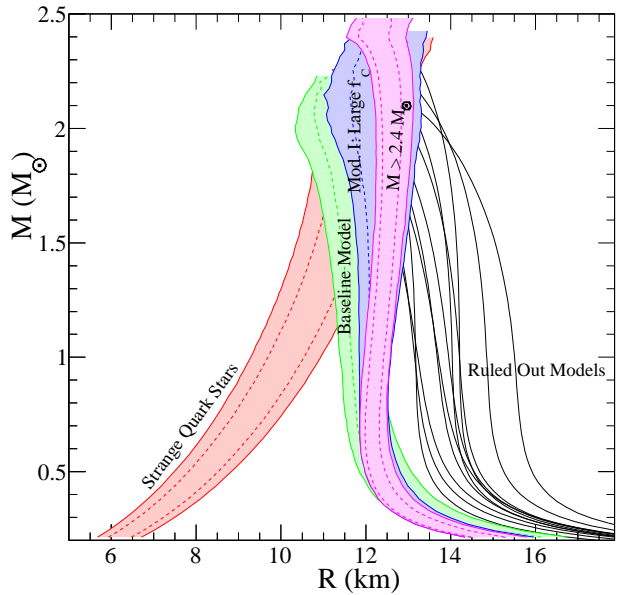}
  \caption{Predicted $M\textrm{--}R$ relations for different EOS models and data
    interpretations. Proceeding from back to front, the red contours
    and probability distributions are for strange quark stars (EOS
    model E with no modifications to the data). Next are green
    contours which correspond to the baseline model (EOS model A with
    no modifications to the data set), and the magenta results are
    those assuming a larger maximum mass to accomodate a mass of 2.4
    solar masses for B1957+20. Finally, the black lines are the 10
    Skyrme models from Stone et al. which are inconsistent with the
    data because their radii are too large (they do not match the
    observations at masses low enough to accomodate the low-mass
    objects like M13).}
  \label{fig:mvsr2}
\end{figure}

While we are able to significantly constrain the $P\textrm{--}\varepsilon$
relation, determination of the composition of neutron star cores is
not yet possible. To this end, we consider EOS model E, which
describes the entire star by the high-density quark matter EOS used in
model D, i.e a self-bound strange quark star. In the mass range 1.4--2
solar masses, the radii are not significantly different from our
baseline model so that there is no strong preference for either
strange quark or hadronic stars; however, model E predicts radii
significantly less than 10 km for low masses
($\le1.2~\mathrm{M}_{\odot}$). 

Our neglect of rotation is unlikely to affect our conclusions.
Rotation increases the radius at the equator and decreases the radius
at the poles, and this could be relevant for the interpretation of
some PRE X-ray bursts: the rotation rate of 4U~1608--522 is 619 Hz,
more than half of the rate for which the equatorial radius
is increased by about 50\%. However, this is likely to produce a
systematic uncertainty smaller than that due to variations in $f_C$,
which we have already taken into account. The rotation rates for the
qLMXBs in our sample are unknown. Assuming they are similar to other
qLMXBs, however, means that the effect of rotation is smaller than
that of their distance uncertainties.

The relationship between pressure and energy density (Figure~2) that
we determine from our baseline analysis from observations is
consistent with effective field theory~\cite{Hebeler10} and quantum
Monte Carlo~\cite{Gandolfi12,Steiner12} calculations of low-density
neutron matter. Note that these neutron matter results are
incompatible with Suleimanov's interpretation of 4U
1724-307~\cite{Suleimanov11} and suggested exclusion of short PRE
bursts and qLMXLBs M13 and $\omega$ Cen. Our results are also
consistent with the high-density constraints from heavy-ion
collisions~\cite{Danielewicz02}.

Our results imply that over one third of the modern Skyrme models
studied in Stone et al.~\cite{Stone03} are inconsistent with
observations. Covariant field-theoretical models that have symmetry
energies which increase nearly linearly with density, such as the
model NL3~\cite{Lalazissis97}, are also inconsistent with our results,
although they may still adequately describe isospin-symmetric matter
in nuclei.

\begin{figure}
\includegraphics[width=3.4in]{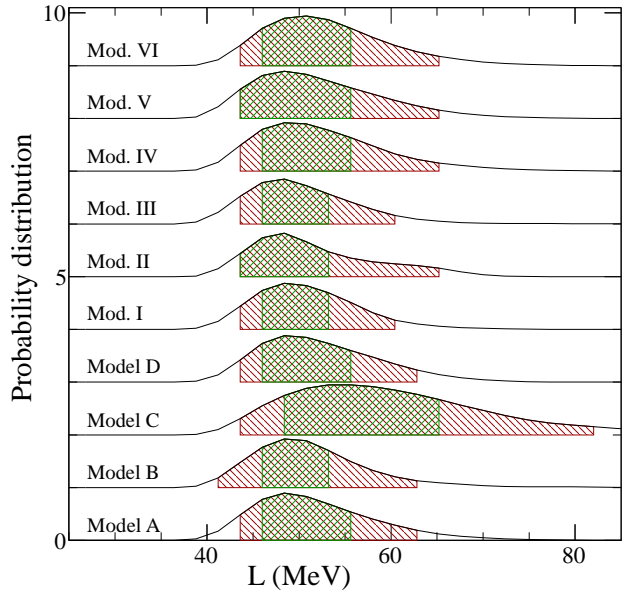}
\caption{The limits on the density derivative of the symmetry energy,
 $L$. The single-hatched (red) regions show the 95\% confidence 
limits and the double-hatched (green) regions show the 68\% 
confidence limits.}
\end{figure}

Our models do not place effective constraints on the symmetry
parameter $S_v$, but do place significant constraints on the symmetry
energy parameter $L$; these are summarized in Figure 4. The
probability distribution for each model is renormalized to fix the
maximum probability at unity and are then shifted upwards by an
arbitrary amount. The range which encloses all of the models and
modifications to the data is 43.3 to 66.5 MeV to 68\% confidence and
41.1 to 83.4 MeV to 95\% confidence. The allowed values of $L$ are
substantially larger for Model C, which allows strong phase transitions
at high densities, because the parametrization decouples the low- and
high-density behaviors.

Our preferred range for $L$ is similar to that obtained from other
astrophysical studies~\cite{Steiner12,Lattimer12,Tsang12} and
experimental studies, e.g., Refs.~\cite{Tsang12,Tamii11}. Our results
suggest that the neutron skin thickness of
$^{208}$Pb~\cite{Typel01,Steiner05} is less than about 0.20 fm. This
is compatible with experiment~~\cite{Horowitz01} and also with
measurements of the dipole polarizability of
$^{208}$Pb~\cite{Reinhard10}.


While we have endeavored to take into account some 
systematic uncertainties in our analysis, we cannot rule out
corrections due to the small number of sources and to possible drastic
modifications of the current understanding of low-mass X-ray binaries.
Nevertheless, it is encouraging that these astrophysical
considerations agree not only with nuclear physics experiments but
also with theoretical studies of neutron matter at low densities and
heavy-ion experiments at higher densities.

We thank G. Bertsch, J. Linnemann and S. Reddy for useful discussions.
This work is supported by Chandra grant TM1-12003X (A.W.S.),
by the Joint Institute for Nuclear Astrophysics at MSU under NSF PHY
grant 08-22648 (A.W.S. and E.F.B.), by NASA ATFP grant NNX08AG76G
(A.W.S. and E.F.B.), and by DOE grants DE-FG02-00ER41132 (A.W.S.) and
DE-AC02-87ER40317 (J.M.L.). J.M.L. and E.F.B thank the Institute for
Nuclear Theory at the University of Washington for partial support
during this work. E.F.B. is a member of an International Team in Space
Science on type I X-ray bursts sponsored by the International Space
Science Institute (ISSI) in Bern, and thanks the ISSI for hospitality
during part of this work.

\bibliographystyle{apsrev2}
\bibliography{paper} 

\end{document}